# Fourier Uniformity: An Useful Tool for Analyzing EEG Signals with An Application to Source Localization

Kaushik Majumdar

*Abstract*—If two signals are phase synchronous then the respective Fourier component at each spectral band should exhibit certain properties. In a pair of artificially generated phase synchronous signals the phase difference at each frequency band changes very slowly over the subsequent frequency bands. This has been called Fourier uniformity in this paper and a measure of it has been proposed. An usefulness of this measure has been outlined in the case of cortical source localization of scalp EEG.

## I. INTRODUCTION

How the EEG signals from different parts of the brain evolve in time with respect to each other is an important criterion to study many physiological (such as cognition) and pathological (such as seizure) brain functions. A key to study this criterion is synchronization, a loosely defined term in neuroscience, biology and even in physics [1], [2]. Since the term does not have a strict definition its measurement also varies from one application to another [3]. In this paper we will primarily be concerned with phase synchronization.

It was Huygens who first studied phase synchronization between two coupled oscillators. A signal can␣be the thought of as a superposition of many coupled oscillators each of which is represented by a Fourier component of the signal. A dominant trend of determining (instantaneous) phase of a signal is with the help of the Hilbert transformation [4], [5]. Another trend is to determine the phase by convolution of the signal with Morlet's wavelet [6], [7]. A Shannon entropy based measure of phase synchronization has also been proposed [8]. The notion of phase in any of these methods is not as natural and readily comprehensible as in Huygens' work. In this paper it has been shown that the notion of phase synchronization between any two signals can be treated from Huygens' point of view. To bridge the gap between Huygens' phase and the phase synchronization between any two arbitrary signals a new notion has been introduced, which is named *Fourier uniformity*. Since the study of synchronization in signals from different parts of the brain is an important area of research, Fourier uniformity may be a potential tool to analyze neural signals, particularly the electrophysiological signals either on the scalp or from the deep brain implants. Apart from describing Fourier uniformity theoretically in this paper we will show its application in cortical source estimation of the scalp human EEG data during median nerve stimulation.

In the next section we will describe the notion of Fourier uniformity. In section 3 we will describe the stimulation experiment, data acquisition and preprocessing. In section 4 we will be presenting the results of our studies first with simulated EEG signals generated from known sources by forward calculation. All the simulations have been done on the real head model of the subject, which has been constructed with the help of his structural MRI data. Then the method has been applied to the EEG signal of the same subject during median nerve stimulation. In the concluding section we will summarize the results with a view to future directions.

## II. FOURIER UNIFORMITY

Any two signals $x_j(t)$ and $x_k(t)$ can be written in terms of their Fourier expansion as

$$x_j(t) = \frac{a_{j0}}{2} + \sum_{n=1}^{\infty} (a_{jn}^2 + b_{jn}^2)^{1/2} \sin\left(\frac{2\pi nt}{p} + \alpha_{jn}\right) \quad (1)$$

$$\alpha_{jn} = \tan^{-1}\left(\frac{a_{jn}}{b_{jn}}\right) \quad (2)$$

$$x_k(t) = \frac{a_{k0}}{2} + \sum_{n=1}^{\infty} (a_{kn}^2 + b_{kn}^2)^{1/2} \sin\left(\frac{2\pi nt}{p} + \alpha_{kn}\right) \quad (3)$$

$$\alpha_{kn} = \tan^{-1}\left(\frac{a_{kn}}{b_{kn}}\right). \quad (4)$$

$n$ is frequency and $p$ is the time duration of the signal. $a_{jn}$

Kaushik Majumdar is with the Center for Complex Systems and Brain Sciences, Florida Atlantic University, 777 Glades Road, Boca Raton, FL 33431, USA, (phone: ++1-561-297-2225; fax: ++1-561-297-3634; e-mail: majumdar@ccs.fau.edu).

and $b_{jn}$ are coefficients of the sin and the cos respectively. Note that the Fourier expansion is valid when the signal is stationary. We are going to use the Fourier expansion for a comparative study between two signals in a way that will not be affected even if the signals are non-stationary as discussed at the end of the next section. By *Fourier uniformity* we mean

$$\alpha_{j1} - \alpha_{k1} = \ldots\ldots = \alpha_{jn} - \alpha_{kn} = \ldots, \quad (5)$$

in strict sense of terms. In other words *Fourier uniformity* between any two signals is the condition where the phase difference between two Fourier components of the same frequency band remains exactly the same across all the bands. However (5) is too strict a condition. To broaden the scope of applicability we propose the following modification in (5):

$$\alpha_{j1} - \alpha_{k1} \approx \ldots\ldots \approx \alpha_{jn} - \alpha_{kn} \approx \ldots, \quad (6)$$

$\approx$ denotes approximately equal. $a \approx b$ is equivalent to $|a - b| < \delta$, where $\delta \to 0$. We will understand satisfaction of (6) for any two signals $x_j(t)$ and $x_k(t)$ as the *approximate Fourier uniformity*.

### III. Phase Synchronization

Following is the standard definition of phase synchronization (for motivation and detailed discussions see [2], [5]).

*Definition 1:* If $\alpha$ and $\beta$ are the phases of two distinct signals or systems then they are *synchronous* if and only if $m\alpha - n\beta = C$, where $C$ is a fixed number such that $0 \leq C < \pi$, where $m, n$ are integers.

For loosely coupled signals or systems a weaker condition is adopted for which $|m\alpha - n\beta| < $ (a constant) holds [2], [5]. We can rewrite this as

$$|m\alpha - n\beta - C| < \delta, \quad (7)$$

where $\delta$ is a small positive quantity. The condition (7) is the satisfiable condition for a pair of systems to be called loosely coupled. For them the notion of synchronization would be replaced by *approximate synchronization*. For convenience of calculation in this paper we shall keep $m = n = 1$. The general case has been dealt in [9].

Usually the phase $\alpha$ of $x_j(t)$ for a given $t$ is calculated with the help of the Hilbert transformation on $x_j(t)$ (for detail see [4], [5]). Phase is also determined for a time window as well as for a frequency window by convolution with a suitable wavelet (for detail see [6], [7]). The notion of phase and their measurements are not same in these two most popular methods.

We have already mentioned that a signal can be viewed as superposition of many coupled oscillators. To elaborate this point consider equation (1). First the raw signal $x_j(t)$ has been collected. To resolve the signal into its principle components integral Fourier transforms for all integers $n$ have been performed. This has yielded the Fourier coefficients $a_{jn}$ and $b_{jn}$ with which the signal can be reconstructed as shown in (1). In reality the Fourier transform is an FFT with only a finite number of terms, which can not fully represent the signal. However the more is the number of terms the better is the representation. Therefore (1) should actually be written as

$$x_j(t) \approx \frac{a_{j0}}{2} + \sum_{n=1}^{\infty}(a_{jn}^2 + b_{jn}^2)^{1/2} \sin\left(\frac{2\pi nt}{p} + \alpha_{jn}\right). \quad (1')$$

However we will assume the sample frequency for digitization of the signal is high (1000 Hz or more) and therefore the approximation is good and we will continue to use (1) instead. In other words $x_j(t)$ has been generated by simultaneous vibration of the harmonic oscillators given by $\left(a_{jn}^2 + b_{jn}^2\right)^{1/2} \sin\left(\frac{2\pi nt}{p} + \alpha_{jn}\right)$ for all natural $n$. Same is the case for $x_k(t)$, which is represented by (3). Now if $x_j(t)$ and $x_k(t)$ are phase synchronous then

$$\left(a_{jn}^2 + b_{jn}^2\right)^{1/2} \sin\left(\frac{2\pi nt}{p} + \alpha_{jn}\right) \quad \text{and}$$

$$\left(a_{kn}^2 + b_{kn}^2\right)^{1/2} \sin\left(\frac{2\pi nt}{p} + \alpha_{kn}\right) \quad \text{must be phase}$$

synchronous. But this does not make much sense, for $\alpha_{jn}$ and $\alpha_{kn}$ are constant and therefore their difference is going to be constant in any way irrespective of the phase synchrony of $x_j(t)$ and $x_k(t)$. The trick of phase synchrony between $x_j(t)$ and $x_k(t)$ lies in the Fourier uniformity between the two signals. In other words they are phase synchronous if and only if (5) holds. Note that unlike [4], [5], [6], [7] we haven't tried to define phase for the whole signal. We have taken the phase of each harmonic component instead, about whose definition there is no dispute whatsoever since the time of Huygens. (5) implies that the phase difference across all the harmonics between the two signals are same and this quantity is the phase difference between the signals. This sounds natural in case of exactly phase synchronous signals. Let us redefine phase synchronization in the following manner.

*Definition 2:* Two signals $x_j(t)$ and $x_k(t)$ are *phase synchronous* or *approximately phase synchronous* if and only if Fourier uniformity or approximate Fourier uniformity holds between them respectively.

Since the coefficients are measured by FFT the sample frequency comes into play and in reality no component becomes zero. So $\alpha_{jn} - \alpha_{kn}$ can be written as $\frac{a_{jn}b_{kn} - a_{kn}b_{jn}}{a_{jn}a_{kn} + b_{jn}b_{kn}}$. If $a_{jn}a_{kn} + b_{jn}b_{kn} = 0$, then by (2) and

$$\tan(\alpha_{jn} - \alpha_{kn}) = \frac{a_{jn}b_{kn} - a_{kn}b_{jn}}{a_{jn}a_{kn} + b_{jn}b_{kn}} = \infty \quad (4)$$

or in other words $\alpha_{jn} - \alpha_{kn} = \frac{\pi}{2}$. If we do not want to take all the frequencies, but want to test the synchronization across specific frequency bands only we can verify the validity of the following equation instead of (5)

$$\alpha_{jn_1} - \alpha_{kn_1} = \alpha_{jn_2} - \alpha_{kn_2} = \ldots = \alpha_{jn_i} - \alpha_{kn_i}, \quad (8)$$

where $n_1, \ldots, n_i$ are the bands to be chosen. However in this paper we will stick to (5) rather than (8).

For wider applicability approximate phase synchronization signified by (6) would be given a measure in this paper. (6) implies

$$\frac{a_{j1}b_{k1} - a_{k1}b_{j1}}{a_{j1}a_{k1} + b_{j1}b_{k1}} \approx \ldots \approx \frac{a_{jn}b_{kn} - a_{kn}b_{jn}}{a_{jn}a_{kn} + b_{jn}b_{kn}} \approx \ldots, \quad (9)$$

implies that both the mean and the standard deviation of $\frac{a_{jn}b_{kn} - a_{kn}b_{jn}}{a_{jn}a_{kn} + b_{jn}b_{kn}} - \frac{a_{jn+1}b_{kn+1} - a_{kn+1}b_{jn+1}}{a_{jn+1}a_{kn+1} + b_{jn+1}b_{kn+1}}$ over all $n$ would be small when $x_j(t)$ and $x_k(t)$ are approximately phase synchronous. Let

$$E(n) = \left| \frac{a_{jn}b_{kn} - a_{kn}b_{jn}}{a_{jn}a_{kn} + b_{jn}b_{kn}} - \frac{a_{jn+1}b_{kn+1} - a_{kn+1}b_{jn+1}}{a_{jn+1}a_{kn+1} + b_{jn+1}b_{kn+1}} \right| \quad (10)$$

A normalized measure of phase synchronization between $x_j(t)$ and $x_k(t)$ is given by the $syn$ function defined as

$$syn(x_j(t), x_k(t)) = syn(j,k) = \frac{1}{1 + mean(E(n)) + std(E(n))}. \quad (11)$$

In case of phase synchronous signals $E(n)$ is a small quantity for all $n$ and therefore both $mean(E(n))$ and $std(E(n))$ are small quantities and $syn(j,k) \approx 1$. When the signals are phase asynchronous $E(n)$ must have to be large for some $n$ and both $mean(E(n))$ and $std(E(n))$ will be large quantities leading to $syn(j,k) \approx 0$. The values of the $syn$ function between 0 and 1 signifies various degrees of phase synchrony between the two signals. Note that $syn$ is purely a measure of phase relationship between two signals and to the best of knowledge of the author has been proposed in this paper only.

Before finishing this section let us emphasize that the relations (6) and (9) will hold for phase synchronous non-stationary signals as well, although Fourier expansion of non-stationary signals may generate significant power distribution at spurious frequency bands. But then if it endows a spurious power to the signal A at frequency n, it will also endow a spurious power to the signal B at n. This is not going to affect the phase relationship between the two signals.

## IV. SIMULATION STUDIES

We have verified the usefulness of (11) on artificially generated EEG signals in the real head model of a 35 year old male volunteer. The head is modeled with the help of his anatomical MRI data. Brain (the cortical surface), skull and scalp have been modeled as nested surfaces each with uniform impedance. Cortical surface has been modeled as a triangular mesh with 8959 points. On scalp there are 60 electrodes whose positions have been recorded by MRI (there were actually 64 electrodes in 10/10 montage [10], data from 4 were not accepted). EEG signals on the scalp have been generated by forward calculation according to Boundary Elements Method (BEM) from a pair of separated sources in the cortical surface. It was done with the help of an open source software called OPENMEEG, developed by the Odyssee group in INRIA Sophia Antipolis and ENPC in France [12]. The detail of the forward calculation is available in [11].

Sources have been modeled as monopoles, which may be thought as dipoles placed normal to the cortical surface whose other ends away from the surface have been ignored. 4 to 7 closely spaced cortical points have been activated by ERP (without additive noise) represented by a periodic signal. The signal has been modeled like $x_j(t)$ but with nonzero $a_{jn}$ and $b_{jn}$ only for 3 arbitrarily chosen $n$, for all other $n$ $a_{jn} = b_{jn} = 0$.

The scalp EEG has been generated by a forward calculation based on the real head model. Boundary elements method has been followed for three surfaces representing cortex, skull and scalp. The detail of the calculation can be found in [11]. Its implementation is available as an open source package called OPENMEEG [12]. When there is only one source after the scalp EEG has been generated synchronization between any two pair of electrodes has been calculated according to (11) and in all cases it has been exactly 1, that is, EEG at any two channels are perfectly phase synchronous. The experiment has been repeated several times by changing the source positions with identical results. This shows (11) can indeed measure perfect phase synchrony.

Next, ten signals have been generated in time domain much the same way the ERP at the cortical source has been generated. One such signal with 400 time points and sampled at 5000 Hz is given below.

$$S_1(t) = 1 + 10\cos(10\pi t) + \sin(10\pi t) + 137\cos(34\pi t) \\ + 9\sin(34\pi t) + 79\cos(194\pi t) + 45\sin(194\pi t)$$

(12)

This way generating phase asynchronous signals is quite straight forward. Out of the 10 only two signals ($S_6(t)$ and $S_7(t)$) are mutually phase synchronous (one has been generated from the other just by suppressing by a factor of 0.5), the rest are all mutually phase asynchronous. Out of the $^{10}C_2 = 45$ values of synchronization 44 values are less than or equal to 0.0881 and $syn(6,7) = 1$.

## V. RESULT

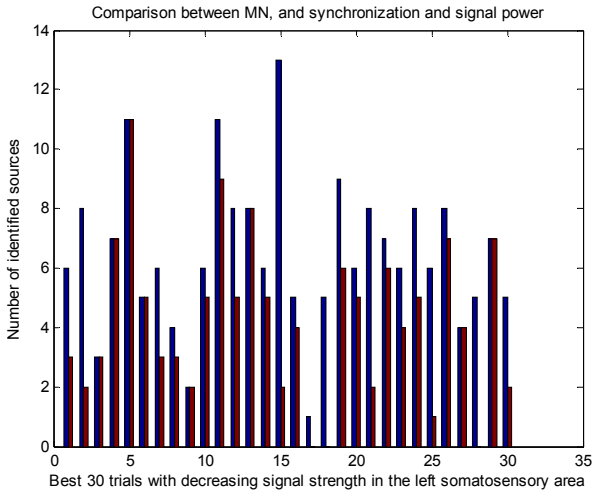

Fig. 1. Number of identified sources (Y-axis) in the 30 best trials (X-axis) have been shown by histogram. 30 best trials have been selected based on the strength of the EEG signals in the somatosensory area. Blue bar (the left portion) indicates sources identified by MN and red (the right portion) indicates those identified by phase synchronization and signal power profile during median nerve stimulation trials.

Human scalp EEG source localization has been performed with the help of (11), the detail of which has been reported in [13]. Cortical source of scalp EEG have been identified within an error of 4.5 cm as measured on the scalp. The location of a cortical source on the scalp has been marked by a channel situated closest to the source. This has been done in two steps. First appropriate neighborhood of each channel has been constructed. Then phase synchronization has been measured by (11) between the channel and each neighbor. Mean of that synchronization value has been calculated. Also the average signal power in the neighborhood has been calculated. Only those channels have been identified as closest to the potential sources where the phase synchronization and signal power within the neighborhood remain above certain threshold. Sources have also been independently identified by the classical minimum norm (MN) inverse method. How they have matched with each other has been shown in Fig. 1.

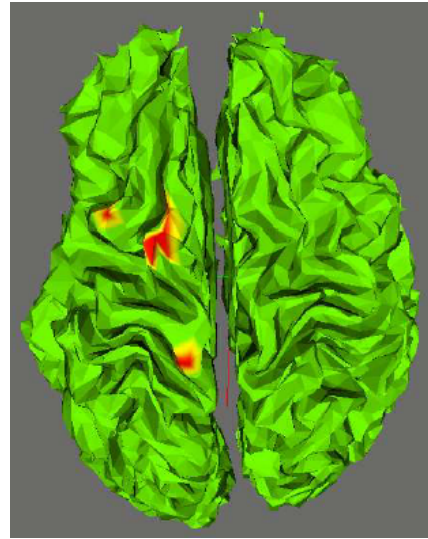

Fig. 2. The sources most active during the best 30 right median nerve stimulation trials during 25 ms after the stimulus onset. For better localization accuracy the sources have been identified independently by (1) MN inverse and (2) the phase synchronization and signal power profile across the scalp EEG channels. Only those sources have been recognized which have been identified by both.

The aim in [13] was to locate the sources in single trials. So suitable trials have been selected. The criterion was having prominent EEG signals in the electrodes covering the somatosensory area, because the trial was median nerve stimulation. Sources have been identified in each of them. The combined result for the 30 trials has been shown in Fig. 2.

The MN inverse has been implemented in the OPENMEEG [12]. The theoretical description of the implementation is available in [14]. Details of the single trial results have been reported in [13].

## VI. CONCLUSION

A comparative study between two signals is an important area of research. Fourier uniformity is the condition in which the phase difference between two Fourier components at a band width remains almost the same across all the band widths. This has been taken as the condition for phase synchrony between the two signals. This notion of phase synchronization has been validated by simulations. An application has been shown in the cortical source localization of the human scalp EEG during median nerve stimulation.

Fourier uniformity is quite sensitive to noise. In the current study we have not included the effects of additive noise to the signals. However even a 1 ms sliding of the time window for the study shows significant change in the Fourier uniformity profile, while the signal power profile remains virtually unchanged. This indicates that an additive noise will significantly alter the Fourier uniformity. But this is true for any other phase synchronization detection method. If a strong noise is added to two weak signals, because of the dominant effect of the same noise the signals may appear phase synchronous, although the original signals may actually be significantly phase asynchronous. If the frequency range of the noise is known (or can be estimated) the Fourier uniformity can be measured across a range outside the frequency range of the noise to ascertain the actual phase synchronization or asynchronization between the signal pair. In other words by selecting the frequency range of Fourier uniformity we can use it as a band-pass filter to remedy the effect of noise. In this paper we have not done this. Instead considered Fourier uniformity across all frequencies.

ACKNOWLEDGMENT

The current work was done when the author was with the Odyssee group in INRIA Sophia Antipolis, France. It was partially supported by the EADS grant 2118.